\documentclass{mn2e}
\usepackage{epsfig}
\def\msun{{\rm M_{\odot}}}

\def\me{{\dot M_{\rm Edd}}}
\def\md{{\dot M_{\rm dyn}}}
\def\mo{{\dot M_{\rm out}}}
\def\le{{L_{\rm Edd}}}
\def\rsh{{R_{\rm shock}}}
\def\rinf{{R_{\rm inf}}}

\title[Black Hole Outflows] {Black Hole Outflows}

\author[A. R. King] {A. R. King$^{1}$ \\
$^1$Theoretical Astrophysics Group, University of Leicester, Leicester
LE1 7RH}

\date{\today}

\volume{000}

\pagerange{\pageref{firstpage}--\pageref{lastpage}} \pubyear{2005}

\begin{document}

\label{firstpage}

\maketitle

\begin{abstract}

I show that Eddington accretion episodes in AGN are likely to produce
winds with velocities $v \sim 0.1c$ and ionization parameters up to
$\xi \sim 10^4$~(cgs), implying the presence of resonance lines of
helium-- and hydrogenlike iron. These properties are direct
consequences of momentum and mass conservation respectively, and agree
with recent X--ray observations of fast outflows from AGN. Because the
wind is significantly subluminal, it can persist long after the AGN is
observed to have become sub--Eddington. The wind creates a strong
cooling shock as it interacts with the interstellar medium of the host
galaxy, and this cooling region may be observable in an inverse
Compton continuum and lower--excitation emission lines associated with
lower velocities. The shell of matter swept up by the
(`momentum--driven') shocked wind must propagate beyond the black
hole's sphere of influence on a timescale $\la 3\times
10^5$~yr. Outside this radius the shell stalls unless the black hole
mass has reached the value $M_{\sigma}$ implied by the $M - \sigma$
relation. If the wind shock did not cool, as suggested here, the
resulting (`energy--driven') outflow would imply a far smaller SMBH
mass than actually observed. In galaxies with large bulges the black
hole may grow somewhat beyond this value, suggesting that the observed
$M -\sigma$ relation may curve upwards at large $M$. Minor accretion
events with small gas fractions can produce galaxy--wide outflows with
velocities significantly exceeding $\sigma$, including fossil outflows
in galaxies where there is little current AGN activity. Some rare
cases may reveal the energy--driven outflows which sweep gas out of
the galaxy and establish the black hole--bulge mass relation. However
these require the quasar to be at the Eddington luminosity.

\end{abstract}


\section{Introduction}

Outflows and winds are widely observed in many types of galaxies, from
active to normal (e.g. Chartas et al., 2003; Pounds et al., 2003a,
2006; O'Brien et al., 2005, Holt et al., 2008; Krongold et al., 2007;
Tremonti et al., 2007). There is often a strong presumption that the
central supermassive black hole (SMBH) is implicated. Although other
types of driving exist, e.g. by supernovae or starbursts, in many
cases these phenomena are themselves associated with accretion
episodes on to the SMBH.

From a theoretical viewpoint, outflows driven by black holes offer a
simple way of establishing relations between the SMBH and its host
galaxy, and hence potential explanations for the $M - \sigma$ and $M -
M_{\rm bulge}$ relations (Ferrarese \& Merritt, 2000; Gebhardt et
al. 2000; H\"aring \& Rix 2004). Such outflows are plausible, as AGN must
feed at high rates to grow the observed SMBH masses, given the short
duty cycle implied by the rarity of AGN among all galaxies. There is
no obvious reason why these rates should respect the hole's Eddington
limit, and outflows are a natural consequence.

However it is unclear just how, if at all, the various types of
observed outflows mentioned in the first paragraph fit together in
terms of these ideas. This paper aims to clarify this.

\section{The Eddington ratio in AGN}

Super--Eddington accretion is widely observed in accreting binary
systems. For example the well--known system SS433 has an Eddington
ratio $\dot m = \dot M/\dot M_{\rm Edd}$ of order 5000 (e.g. King et
al., 2000; Begelman et al., 2006). Here $\dot M, \dot M_{\rm Edd}$ are
the accretion rate and the Eddington value respectively.  

The distinctive feature of AGN accretion is that such large Eddington
ratios are unlikely. To see this we note that the maximum possible
accretion rate is the dynamical rate
\begin{equation}
\dot M_{\rm dyn} \simeq {f_g \sigma^3\over 2G},
\label{dyn}
\end{equation}
The dynamical rate applies when a gas mass which was previously
self--supporting against gravity is destabilized and falls freely on
to the black hole. Equation (\ref{dyn}) describes the case where gas
is initially in rough virial equilibrium in the bulge of a galaxy with
velocity dispersion $\sigma$ and baryonic mass fraction
$f_g$. Parametrizing, we find
\begin{equation}
\md \simeq 1.4\times 10^2 \sigma_{200}^3~\msun\, {\rm yr}^{-1}
\label{dyn2}
\end{equation}
where $\sigma_{200} = \sigma/(200~{\rm km\, s^{-1}})$, and we have taken
$f_g = 0.16$. We compare this with
\begin{equation}
\me = {\le\over \eta c^2} = {4\pi GM\over \kappa\eta c}
\label{edd}
\end{equation}
where $\le$ is the Eddington luminosity, $\eta$ the radiative
efficiency of accretion, and $\kappa$ the electron scattering
opacity. We evaluate this for $\eta = 0.1$ and black hole masses $M$
lying close to the observed $M - \sigma$ relation
\begin{equation}
M \simeq 2\times 10^8\msun\sigma_{200}^4
\label{msig}
\end{equation}
to find
\begin{equation}
\me \simeq 4.4~\sigma_{200}^4\msun\, {\rm yr}^{-1}
\end{equation}
and thus an Eddington ratio 
\begin{equation}
\dot m < {\md \over \me} \simeq {33\over \sigma_{200}} \simeq
     {39\over M_8^{1/4}}
\label{eddrat}
\end{equation}
where $M_8 = M/10^8\msun$. Since $0.1 \la M_8 \la 10$ for the black holes in
AGN, and $\md$ is an upper limit to $\dot M$, modest values $\dot m \sim 1$ of
the Eddington ratio are likely.

For completeness I note that much higher Eddington ratios can occur
if the gas mass which is destabilized was previously held together by
self--gravity, e.g. as in a star. In this case we find a maximum
dynamical rate $\md \sim v_{\rm orb}^3/2G \sim M_*/P$, where $M_*$ is
the stellar mass and $v_{\rm orb}, P$ its orbital velocity and
period. This gives rates $\md \sim 0.1\msun/{\rm hr}$ in stellar--mass
binary systems. Rather lower values result for tidal disruption of
stars near SMBH because the stellar debris is only accreted over a
spread in orbital times (e.g. Lodato et al., 2009).

\section{SMBH winds at modest Eddington ratios}

Motivated by the results of the previous section, I examine the
properties of winds from SMBH accreting with modest Eddington ratios
$\dot m \sim 1$. I assume that the winds are quasi--spherical, over
solid angle $4\pi b$, with $b \sim 1$ (this is confirmed for
PG1211+143; Pounds \& Reeves, 2009).  It is well known that winds of
this type have electron scattering optical depth $\tau \sim 1$
measured from infinity in to a distance of order the Schwarzschild
radius $R_s = 2GM/c^2$ (e.g. King \& Pounds, 2003). This means that on
average every emitted photon scatters about once before escaping to
infinity, which in turn suggests that the total wind momentum must be
of order the photon momentum, i.e.
\begin{equation}
\mo v \simeq {\le\over c},
\label{mom}
\end{equation}
as is for example also found for the winds of hot stars. Using
(\ref{edd}) gives the wind velocity
\begin{equation}
v \simeq {\eta\over \dot m}c \sim 0.1c.
\label{v}
\end{equation}
This agrees with the fact that winds are always found to have a
terminal velocity typical of the escape velocity from the radius at
which they are launched, here several tens of $R_s$. As expected, using
this value of $v$ in the mass conservation equation
\begin{equation}
\mo = 4\pi bR^2v\rho(R),
\label{mass}
\end{equation}
where $\rho(R)$ is the mass density, self--consistently shows that the
optical depth $\tau$ of the wind is $\sim 1$ (cf King \& Pounds,
2003, eqn. 4). 

Since the wind moves with speed $\sim 0.1c$, it can persist long after
the AGN is observed to have become sub--Eddington. The duration of the
lag is $\sim 10R/c$, where $R$ is the radial extent of the wind
(i.e. the shock radius, as we shall see below). For $R \ga 3$~pc this
lag is at least a century, and far longer lags are possible, as we
shall see. This may be the reason why AGN showing other signs of
super--Eddington phenomena (e.g. narrow--line Seyfert 2 galaxies) are
nevertheless seen to have sub--Eddington luminosities (e.g. NGC 4051:
Denney et al., 2009).

We can use (\ref{v}, \ref{mass}) to estimate the ionization
parameter 
\begin{equation} 
\xi = {L_i\over NR^2}
\label{ion}
\end{equation}
of the wind. Here $L_i = l_i\le$ is the ionizing luminosity, with
$l_i< 1$ a dimensionless parameter specified by the quasar spectrum,
and $N = \rho/\mu m_p$ is the number density. This gives
\begin{equation}
\xi = 3\times 10^4\eta_{0.1}^2l_2\dot m^{-2},
\label{ion2}
\end{equation}
where $l_2 = l_i/10^{-2}$, and $\eta_{0.1} = \eta/0.1$. 

Equation (\ref{ion2}) shows that the wind momentum and mass rates
determine its ionization parameter: for a given quasar spectrum, the
predominant ionization state is such that the threshold photon energy
defining $L_i$, and the corresponding ionization parameter $\xi$,
together satisfy (\ref{ion2}). This requires high excitation: a low
threshold photon energy (say in the infrared) would imply a large
value of $l_2$, but the high value of $\xi$ then given by (\ref{ion2})
would require the presence of very highly ionized species, physically
incompatible with such low excitation.

For suitably chosen continuum spectra, it is possible to envisage a
range of solutions of (\ref{ion2}), and it is even possible that a
given spectrum may allow more than one solution, the result being
specified by initial conditions. However for a typical quasar
spectrum, an obvious self--consistent solution of (\ref{ion2}) is $l_2
\simeq 1$, $\dot m \simeq 1$, $\xi \simeq 3\times 10^4$.  This
describes the case where the quasar is currently radiating at the
Eddington limit. However as remarked after eqn (\ref{mass}), we can
also have situations where the quasar's luminosity has dropped after
an Eddington episode, but the wind is still flowing, with $\dot m
\simeq 1$. In this case the ionizing luminosity $10^{-2}l_2\le$ in
(\ref{ion2}) takes a lower value, giving a lower value of $\xi$. For
example a quasar of luminosity $0.3\le$ would have $\xi \sim 10^4$.
This corresponds to a photon energy threshold appropriate for helium--
or hydrogenlike iron (i.e. $h\nu_{\rm threshold}\sim 9$~keV).

We conclude that

{\it Eddington winds from AGN are likely to have velocities $\sim
  0.1c$, and show the presence of helium-- or hydrogenlike iron.}

A number of such winds are known (see Cappi, 2006, for a review). This
Section shows that it is no coincidence that in all cases the wind
velocity is $v \sim 0.1c$, and further that they are all found by
identifying blueshifted resonance lines of Fe XXV, XXVI in
absorption. Conversely, any observed wind with these properties
automatically satisfies the momentum and mass relations (\ref{mom},
\ref{mass}), strongly suggesting launching by an AGN accreting at a
slightly super--Eddington rate.

Two further remarks are relevant here. We note first that (\ref{mom})
implies a kinetic energy rate
\begin{equation}
{1\over 2}\mo v^2 \simeq {v\over c}\le \simeq {\eta\over 2}\le \simeq 0.05\le
\label{kin}
\end{equation}
implying a mechanical `energy efficiency' $\eta/2\simeq 0.05$ wrt
$\le$. Cosmological simulations typically adopt such values in order
to produce an $M-\sigma$ relation in agreement with observation
(e.g. di Matteo, 2005). As we have seen, this implicitly means that
they adopt the single--scattering momentum relation (\ref{mom}). We
shall see below (in Section 5) that there must also be an implicit
assumption of momentum rather than energy driving, i.e. that the wind
interacts with the host galaxy through its ram pressure rather than
its energy.

I note that the near--Eddington regime considered here differs from
the hyper--Eddington regime discussed by Shakura \& Sunyaev (1973),
where the much larger optical depth leads to the energy relation $\mo
v^2/2 \simeq \le$ instead of the momentum relation (\ref{mom}). This
generally only holds in accreting stellar--mass binary systems (see
Section 6 below), but King (2009) discusses a possible case involving
supermassive black holes.

Finally, we should also note that although we assume here that
Eddington or super--Eddington accretion leads to winds and outflow,
some authors have considered alternative possibilities. The main
observational constraint on these ideas is the Soltan argument
(Soltan, 1982). This relates the background radiation of the Universe
to the mass density in supermassive black holes, and with subsequent
work (e.g. Yu \& Tremaine, 2002) is generally interpreted as implying
that most of the mass of the SMBH in galaxy centres results from
luminous accretion with a typical radiative efficiency $\eta \sim
0.1$. 

However since the value of $\eta$ appearing in the Soltan argument is
a global average over many different accretion events, it is clearly
still possible for some fraction of the mass growth to occur in other
ways which produce little outflow, as for example in the so--called
Polish doughnuts (see Abramowicz, 2009, for a recent review). This
type of accretion flow could reduce or even remove the mechanical
feedback between black hole growth and the host galaxy. The fact that
the $M - \sigma$ relation emerges (without free parameter) from
consideration of momentum--driven Eddington winds tends to support the
straightforward interpretation of the Soltan argument adopted here,
but one should be alive to alternative possibilities.

\section{Interaction with the host}

It is clear that an Eddington wind of the type discussed above can
have a significant effect on its host galaxy. The kinetic power of
the wind is 
\begin{equation}
\mo {v^2\over 2} = {v\over 2c}\le \simeq 0.05\le
\label{power}
\end{equation}
where we have used (\ref{mom}) and (\ref{v}). If the wind persists as the hole
doubles its mass (i.e. for a Salpeter time), its total energy is $\simeq
5\times 10^{59}M_8$~erg, where $M_8$ is the hole mass in units of
$10^8\msun$. This formally exceeds the binding energy $\sim M_{\rm
bulge}\sigma^2 \sim 3\times 10^{58}$~erg of a galaxy bulge with
baryonic mass $M_{\rm bulge} \sim 10^{11}\msun$ and velocity dispersion
$\sigma \sim 200~{\rm km\, s^{-1}}$ (as suggested by the $M - M_{\rm bulge}$
and $M - \sigma$ relations). Evidently the coupling of wind energy to the
galaxy must be inefficient, as black holes would destroy or at least severely
modify their host bulges in any significant super--Eddington growth phase.

To investigate this we must consider how the wind interacts with the
interstellar medium of the host. Just as in the corresponding problem for a
stellar wind, the interaction must successively involve an inner (reverse)
shock, slowing the central wind, a contact discontinuity between the shocked
wind and the shocked, swept--up interstellar medium, and an outer (forward)
shock driven into this medium and sweeping it outwards, ahead of the shocked
wind (see Fig. 1). 
\begin{figure*}

\centerline{\psfig{file=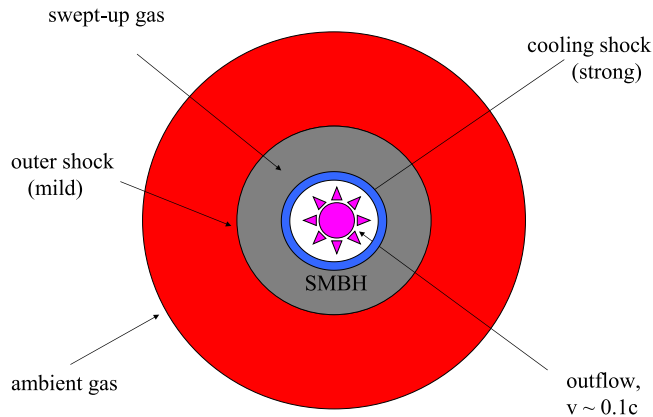,width=0.5\textwidth,angle=0}}
\caption{Schematic view of the shock pattern resulting from the impact
  of an Eddington wind on the interstellar gas of the host galaxy. A
  supermassive black hole (SMBH) accreting at just above the Eddington
  rate drives a fast wind (velocity $u = v \sim \eta c \sim 0.1c$),
  whose ionization state makes it observable in X--ray absorption
  lines. The outflow collides with the ambient gas in the host galaxy
  and is slowed in a strong shock. The inverse Compton effect from the
  quasar's radiation field rapidly cools the shocked gas, removing its
  thermal energy and strongly compressing and slowing it over a very
  short radial extent. This gas may be observable in an inverse
  Compton continuum and lower--excitation emission lines associated
  with lower velocities. The cooled gas exerts the preshock ram
  pressure on the galaxy's interstellar gas and sweeps it up into a
  thick shell (`snowplough'). This shell's motion drives a milder
  outward shock into the ambient interstellar medium. This shock
  ultimately stalls unless the SMBH mass has reached the value
  $M{_\sigma}$ satisfying the $M - \sigma$ relation.}
\label{}
\end{figure*}
The inefficient coupling of wind energy to the galactic baryons noted
above strongly suggests that the shocked wind cools rapidly after
passing through the inner shock. This removes the thermal pressure
generated in the shock, and leaves only the preshock ram pressure
acting on the interstellar medium.
\begin{figure*}

\centerline{\psfig{file=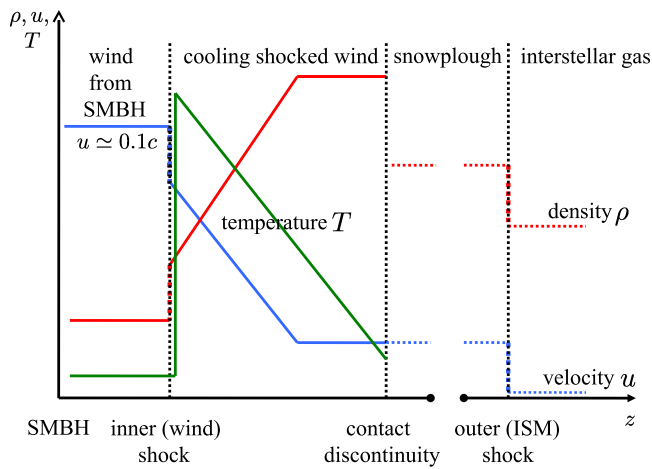,width=0.5\textwidth,angle=0}}
\caption{Impact of a wind from an SMBH accreting at a super--Eddington
  rate on the interstellar gas of the host galaxy: schematic view of
  the radial dependence of the gas density $\rho$, velocity $u$ and
  temperature $T$. At the inner shock, the gas temperature rises
  strongly, while the wind density and velocity respectively increase
  (decrease) by factors $\sim 4$. Immediately outside this (adiabatic)
  shock, the strong Compton cooling effect of the quasar radiation
  severely reduces the temperature, and slows and compresses the wind
  gas still further. This cooling region is very narrow compared with
  the shock radius (see Fig. 1), and may be observable through the and
  inverse Compton continuum and lower--excitation emission lines. The
  shocked wind sweeps up the host ISM as a `snowplough'. This is much
  more extended than the cooling region (cf Fig.1), and itself drives
  an outer shock into the ambient ISM of the host. Linestyles: red,
  solid: wind gas density $\rho$; red, dotted: ISM gas density $\rho$;
  blue, solid, wind gas velocity $u$; blue, dotted, ISM gas velocity
  $u$; green, solid, wind gas temperature $T$. The vertical dashed
  lines denote the three discontinuities, inner shock, contact
  discontinuity and outer shock.}
\label{}
\end{figure*}
A clear candidate for this shock cooling is the inverse Compton effect
on the quasar's radiation field (King, 2003). We note that this
typically has Compton temperature $T_c\sim 10^7$~K, whereas the formal
temperature at the inner adiabatic shock is $m_pv^2/k \sim
10^{11}$~K. The quasar radiation cools the inner shock efficiently,
provided that this is within galaxy--scale distances from the centre
(King, 2003). Inverse Compton cooling should produce a component in
the quasar spectrum characterized by $kT_c \sim 1$~keV and with a
luminosity $\sim \mo v^2/2 \simeq 0.05\le$, i.e. about $5\%$ of the
quasar's bolometric output. I note that even if the quasar becomes
sub--Eddington, leaving a wind persisting for a lag time $10 R_{\rm
  shock}/c$, its radiation field is still able to cool the shock
efficiently.

Given this efficient cooling, the inner (adiabatic) shock and very
narrow associated cooling region together constitute an `isothermal
shock'. The gas density jumps by a factor $\sim 4$ at the adiabatic
shock, accompanied by a velocity drop by the same factor. It is then
strongly compressed in the cooling region while the velocity slows to
low values (see Fig. 2). Since the cooling is efficient the whole
region is very thin compared with the shock radius $R_{\rm shock}$, and
we can regard the shock as locally plane. The Rankine--Hugoniot
relations across this isothermal shock then show that the mass flow
rate $\rho v$ remains constant, while the postshock gas pressure drops
to the value
\begin{equation}
P_{\rm ram} = \rho v^2 = {\dot Mv\over 4\pi bR_{\rm shock}^2} \simeq {\le \over
  4\pi bR_{\rm shock}^2c},
\label{ram}
\end{equation}
i.e. the preshock ram pressure. With a constant cooling time, as
expected, the postshock temperature and velocity $u$ drop
approximately linearly with distance behind the shock, and the density
rises as $1/u$, strongly increasing its emission measure. The gas is
likely to be in photoionization equilibrium as it has low optical
depth to the quasar radiation, and the increased densities imply short
recombination times. The mass conservation equation (\ref{mass}) and
ionization parameter (\ref{ion}) combine to give
\begin{equation}
{l_i u \over \xi} = {\rm constant}
\label{excit}
\end{equation}
in this region. {\it We thus expect a correlation between velocity and
  excitation.} The rapid cooling in this region implies a rapid
transition between the immediate postshock regime ($\sim v/4$, keV
excitation) and the much slower and cooler compressed state. There is
direct observational evidence for this cooling shock in NGC 4051
(Pounds et al., in prep). Pounds et al (2004) had already noted a
correlation of outflow velocity with ionization in this source.

\section{Dynamics}

Given the basic structure sketched in the last Section, we can
investigate how the shock pattern moves through the interstellar
medium of the host galaxy. The cooled postshock gas exerts the ram
pressure (\ref{ram}) on the undisturbed interstellar medium of the
galaxy, driving an outer shock into it and sweeping it up into a
relatively dense shell of increasing mass. The equation of motion of
the shell in the momentum--driven limit is
\begin{equation}
{{\rm d}\over {\rm d}t}[M(R)\dot R] + {GM(R)[M + M_{\rm tot}(R)]\over R^2} = 
4\pi\rho v^2 = {\le\over c}
\label{motion}
\end{equation}
where 
\begin{equation}
M(R) = 4\pi\int_0^R\rho_{\rm ISM} r^2 {\rm d}r
\label{m}
\end{equation}
is the swept--up interstellar gas mass, $M$ is the black hole mass,
$M_{\rm tot}= M(R)/f_g$ is the total mass within radius $R$ (including
any dark matter), and $f_g$ is the gas fraction (note that in eqn (2)
of King, 2005 the suffix `tot' was inadvertently missed off the
relevant quantity). This equation takes different forms depending on
which part of the host galaxy the shell has reached.

\subsection{Close to the black hole}

Close to the black hole, i.e. within its sphere of influence, of radius
\begin{equation}
R_{\rm inf} \simeq {GM\over \sigma^2} \simeq 17M_8\sigma_{200}^{-2}~{\rm pc},
\label{rinf}
\end{equation}
the black hole dominates the gravitational potential, and there is
essentially no dark matter. Then (\ref{motion}) becomes
\begin{equation}
{{\rm d}\over {\rm d}t}[M(R)\dot R] + {GM(R)M\over R^2} = {\le\over c} 
\end{equation}
Multiplying through by $M(R)\dot R/GM$ we find the first integral
\begin{equation}
{[M(R)\dot R]^2\over 2GM} = {4\pi\over \kappa}\int M(R){\rm d}R -
\int{M^2(R)\over R^2} {\rm d}R.
\label{int}
\end{equation}
This equation shows that for any reasonable distribution of matter
$M(R)$, the shell cannot move outwards unless
\begin{equation}
M(R) \la {4\pi R^2\over \kappa} \sim 2\times 10^{-4}\left({R\over
  R_s}\right)^2M_8\msun,
\label{limit}
\end{equation}
where we have parametrized the radius in units of the Schwarzschild
radius $R_s$ of the black hole, with $M_8 = M/10^8\msun$. The physical
content of (\ref{limit}) is that the Eddington thrust cannot lift the
weight of a more massive shell at the radius $R$. An equivalent
formulation is 
\begin{equation}
\tau = {M(R)\kappa\over 4\pi R^2} \la 1
\label{limit2}
\end{equation}
i.e. that the maximum shell mass at a given radius has Thomson depth
$\sim 1$.

We see that even relatively small amounts of gas sufficiently close to
the black hole can stall the outflow. However in this case, the gas
from the central Eddington wind would accumulate at the stalled shock
until its own mass violated the limit (\ref{limit}). The equivalent
(\ref{limit2}) shows that the inner wind shock would become optically
thick to the quasar radiation, causing multiple scattering and
enhancing the momentum deposition. The postshock pressure would begin
to exceed $P_{\rm ram}$ by large factors. Unless the black hole had a
very low mass, this enhanced pressure would cause the shell to move
out again. This argument shows that the shell moves so as to keep the
optical depth of the Eddington wind $\la 1$, i.e.
\begin{equation}
{\mo t\kappa\over 4\pi R^2} \la 1
\end{equation}
so that
\begin{equation}
R \ga \left({\mo\kappa \over 4\pi}t\right)^{1/2} = 5\times
10^{16}\dot m^{1/2}M_8^{1/2} t_{\rm yr}^{1/2}~{\rm cm}
\end{equation} 
and
\begin{equation}
\dot R \ga 1250\dot m^{1/2}M_8^{1/2}t^{-1/2}_{\rm yr}~{\rm km\,
  s^{-1}}
\end{equation}
This implies that the shell would reach a radius $\sim 10^3R_s$ in
no more than a year. At this point the swept--up mass (from (\ref{limit})
could be as large as $200m_8\msun$, which would imply an emission
measure comparable with an AGN broad--line region. The shell would reach
$\sim R_{\rm inf}$ in $\la 3\times 10^5$~yr, with a velocity $\dot R
\ga 2~{\rm km\, s^{-1}}$.

Thus even a shell whose progress is blocked by an unfavourable matter
distribution reaches $R_{\rm inf}$ in about $10^5$~yr, assuming that
the quasar wind continues to drive it. In the opposite extreme, where
the mass of the swept--up matter is low, the time to emerge decreases
as $t \sim M_g^{1/2}$, and is limited only by the wind--travel time
$\sim 10R/c \sim 500$~yr for arbitrarily low $M_g$.

This mechanism is clearly limited to the inner parts of a galaxy
bulge, as a shell driven in this way can typically only reach radii
$\la 200$~pc in a Salpeter time, and $\la 2$~kpc even after a Hubble
time. Accordingly we do not consider this mechanism in the next
subsection, which treats outflows at radii $>\rinf$.

\subsection{Far from the black hole}

Far from the black hole (i.e. for $R > R_{\rm inf}$) the dark matter
term $M_{\rm tot}$ becomes dominant in the equation of motion
(\ref{motion}), and we can drop the black hole mass term involving
$M$. The condition that the shell should just be able to escape to
infinity specifies a relation between the black hole mass $M$ and the
parameters of the galaxy potential, particularly the velocity
dispersion, i.e. an $M - \sigma$ relation. For a general mass
distribution $M_{\rm tot}(R)$ we can use the first integral
(\ref{int}) to do this. However for a simple isothermal potential the
equation of motion has the analytic solution
\begin{equation}
\rsh^2 = \left[{G\le\over 2f_g\sigma^2c} - 2(1-f_g)\sigma^2\right]t^2
+ 2R_0v_0t + R_0^2
\label{shell}
\end{equation}
where $R_0, v_0$ are the position and speed of the shell at time $t=0$ (King,
2005). For large times the first term dominates, and the shell can reach
arbitrarily large radii if and only if the black hole mass exceeds the
critical value
\begin{equation}
M_{\sigma} = {f_g(1-f_g)\kappa\over \pi G^2}\sigma^4 \simeq
{f_g\kappa\over \pi G^2}\sigma^4.
\label{msig2}
\end{equation}
This is very close to the observed $M - \sigma$ relation (\ref{msig})
(cf King, 2005). At sufficiently large radii the quasar radiation
field is too dilute to cool the wind shock, and the shell accelerates
beyond the escape value, cutting off the galaxy and establishing the
black--hole mass -- bulge--mass relation (cf King, 2003, 2005). We see
that the time for a continuously--driven shell to reach a given radius
$R = 10R_{10}$~kpc is
\begin{equation}
t \simeq {R\over \surd{2}\sigma(M/M_{\sigma} - 1)} \simeq 3.5\times
10^7{R_{10}\over \sigma_{200}(M/M_{\sigma} - 1)}~{\rm yr}.
\label{flowtime}
\end{equation}
For large $R_{10}$ this time is significantly longer than the Salpeter
time, implying that the black hole mass $M$ must increase above the
threshold value $M_{\sigma}$ before the shell reaches large
radii. This may suggest that for galaxies with large bulge radii, the
black hole mass may tend to lie above the $M- \sigma$ relation. There
is some suggestion of this in the observational data (Marconi \& Hunt,
2003). However the uncertainty here is in knowing just what radius the
shell must reach in order to shut off further accretion on to the
black hole.

\section{Energy--Driven Outflows}

We see from the reasoning of the last Section that the interaction
beween the quasar wind and its host establishing the $M - \sigma$
relation is -- crucially -- `momentum--driven' rather than
`energy--driven'. This equivalent to requiring efficient shock
cooling. An energy--driven shock (e.g. Silk \& Rees, 1998) would
result in a much smaller black hole mass for for a given $\sigma$ than
observed. Instead of the momentum rate $\le/c$ balancing the weight of
swept--up gas $4f_g\sigma^4/G$, which is what produces the
momentum--driven relation (\ref{msig2}), an energy--driven shock would
equate the energy deposition rate to the rate of working against this
weight. In the near--Eddington regime the result is
\begin{equation}
{1\over 2}\mo v^2 \simeq {\eta\over 2}\le = 2{f_g\sigma^4\over G}.\sigma
\label{work}
\end{equation}
i.e.
\begin{equation}
M({\rm energy}) \simeq {2f_g\kappa\over \eta\pi G^2 c}\sigma^5 =
{2\sigma\over \eta c}M_{\sigma} = 3\times 10^6\msun\sigma^5_{200},
\label{energysig}
\end{equation}
which lies well below the observed relation (\ref{msig}). The coupling
adopted in cosmological simulations evidently ensure that the
interstellar medium feels the outflow momentum rather than its energy,
in addition to the `energy efficiency' $\sim \eta/2 \simeq 0.05$ noted
above.

I note finally that if instead of the near--Eddington regime
considered here, an energy--driven outflow had a large Eddington ratio
$\dot m >>1$, we would have to replace (\ref{work}) by 
\begin{equation}
{1\over 2}\mo v^2 \simeq \le = 2{f_g\sigma^4\over G}.\sigma
\label{work2}
\end{equation}
and thus (\ref{energysig}) by
\begin{equation}
M({\rm energy},\, \dot m >>1) 
= {\sigma\over c}M_{\sigma} = 1.5\times 10^5\msun\sigma^5_{200},
\label{energysig2}
\end{equation}
which is still smaller. Indeed one could imagine a situation in which
the central black holes of medium or large galaxies obeyed
(\ref{energysig2}) rather than the observed (\ref{msig}) and
self--consistently had central accretion rates well above Eddington,
since (\ref{eddrat}) would now become
\begin{equation}
\dot m < {\md \over \me} \simeq {4.4\times 10^4\over \sigma_{200}} \simeq
     {5.2\times 10^4\over M_5^{1/4}}.
\label{eddrat2}
\end{equation}
The high optical depth implied by the large Eddington ratio might
then prevent efficient Compton cooling, justifying the original
hypothesis of energy--driven outflow.

It is interesting that observation does not seem to give examples of
this possibility, i.e. medium or large galaxies with very low--mass
central black holes which could accrete at very high Eddington ratios.
The reason may be the inherent tendency of this low--mass sequence to
move irreversibly over time to the high--mass case specified by the
usual $M - \sigma$ relation (\ref{msig}): thus steady sub-- or
near--Eddington accretion on to such a hole could eventually increase
its mass to the point that it was unlikely to accrete at high
Eddington ratios. This would then imply efficient cooling of the shock and
thus a momentum--driven outflow.

\section{Galaxy--wide high--velocity outflows}

On large scales the outflows described in Section 5.2 above all have (outer)
shock velocities limited by the bulge velocity dispersion $\sigma$. Yet
optical and UV observations give clear evidence of outflows with velocities of
several times this value. These cannot be the central quasar winds with $v
\sim 0.1c$ discussed in Section 3. Outflows confined within a few parsecs of
the black hole may be the near--zone winds inside $\rinf$ discussed in Section
5.1, but high--velocity outflows are often seen or inferred on scales
comparable with the entire galaxy.

Some of these outflows are seen in compact radio sources (e.g. Holt et
al., 2008), some in Seyfert galaxies which are clearly sub--Eddington,
and others in post--starburst galaxies (cf Tremonti et al., 2007). The
latter could result from the combined effects of stellar winds and
supernovae, but the known association of starbursts and AGN leave open
the possibility that black holes may be the ultimate driver.

\subsection{Outflows from minor accretion events}

There is a simple interpretation of such large--scale high--velocity
outflows. Consider a galaxy in which the SMBH has reached the mass
$M_{\sigma}$ given by eqn (\ref{msig2}), with the cosmic gas fraction
$f_g \simeq 0.16$. Its bulge gas will probably be severely
depleted. In a subsequent minor accretion event triggering AGN
activity, the effective gas fraction in the bulge will be $f'_g <
f_g$. If accretion on to the SMBH becomes super--Eddington for a time
$\ga 10^5$~yr, the AGN must drive an outflow shock beyond the radius
$\rinf$. However because of the discrepancy between $f_g$
(establishing the black hole mass), and $f'_g$ (the current gas
fraction), the shell radius now obeys a modified form of the analytic
solution (\ref{shell}), namely
\begin{equation}
\rsh^2 = \left[{G\le\over 2f'_g\sigma^2c} - 2(1-f'_g)\sigma^2\right]t^2
+ 2R_0v_0t + R_0^2
\label{shell2}
\end{equation}
where the $\le$ term involves $f_g$ rather than $f'_g$. Thus at large
$t$ we have
\begin{equation}
\rsh^2 = 2\left[{f_g\over f'_g}(1-f_g) - (1-f'_g)\right]\sigma^2t^2
\simeq 2{f_g\over f'_g}\sigma^2t^2 
\label{shell3}
\end{equation}
where we have taken $f'_g << f_g < 1$ in the last form. This shows
that the shell reaches velocities
\begin{equation}
\simeq
(2f_g/f'_g)^{1/2}\sigma > \sigma, 
\label{vel}
\end{equation}
because its inertia is much lower than the one previously expelled by
the Eddington thrust in the accretion episode which defined the SMBH
mass. If at some point the AGN activity turns off, we can match
another solution of the form (\ref{shell}), but with $\le$ formally $=
0$, to the solution (\ref{shell2}). This solution reveals that a
coasting shell stalls only at distances $\sim (f_g/f'_g)^{1/2}$ times
its radius $R_0$ at the point when AGN activity ceased, and thus
persists for a timescale $R_0/\sigma \sim 10^8$~yr, (cf
\ref{flowtime}).

Episodic minor accretion events of this type therefore naturally produce
large--scale outflows with velocities $> \sigma$. Moreover, since they
persist as fossil winds long after the AGN has become faint, they can
have total momentum considerably higher than could be driven by the
{\it current} AGN radiation pressure, i.e. well in excess of the
apparent momentum limit.

\subsection{Outflows from major accretion events}

There is a second possible type of rapid outflow on large spatial
scales. It is inherently rarer than those triggered by minor accretion
events, as described in the previous subsection, but represents a
decisive stage in the evolution of the galaxy. 

Consider a major Eddington accretion event in which the central SMBH
attains its critical mass $M_{\sigma}$. Equation (\ref{msig2}) shows
that the resulting momentum--driven outflow does not stall, as it does
for lower SMBH masses, but continues out to large radii. Beyond a
critical radius, the quasar radiation field is too dilute to cool the
wind shock on its (momentum--driven) flow time $t_{\rm flow}$. If
Eddington accretion on to the central SMBH is still continuing, the
outflow becomes energy--driven. The extra thermal energy now
accelerates the shell further, eventually to a terminal speed
\begin{equation}
v_e \simeq \biggl[{2\eta\sigma^2c\over 3}\biggr]^{1/3} \simeq
875\sigma_{200}^{2/3}~{\rm km\ s}^{-1}
\label{ve}
\end{equation}
(King, 2005). This is fast enough to drive gas out of the galaxy
potential entirely. 

This process ultimately limits the gas content of the galaxy bulge and
hence establishes the black hole--bulge mass relation (King, 2003,
2005). It is clearly a rarer type of event than outflows driven by minor
accretion events as described above. A clear observational distinction
is that these rare energy--driven outflows require that the
quasar driving the outflow should be radiating at the Eddington limit.

\section{Conclusion}

We have shown that Eddington accretion episodes in AGN produce winds
with velocities $v \sim 0.1c$ and ionization parameters $\xi \sim
10^4$~(cgs) requiring the presence of resonance lines of helium-- and
hydrogenlike iron. These properties follow from momentum and mass
conservation, and agree with recent X--ray observations of high--speed
outflows from AGN. Because the winds have speeds $\sim 0.1c$ they can
persist long after the AGN have become sub--Eddington.

The wind creates a strong cooling shock as it impacts the interstellar
medium of the host galaxy. This cooling region may be observable, as
it produces an inverse Compton continuum and lower--excitation
emission lines associated with lower velocities. The shell of matter
swept up by the ('momentum--driven') shocked wind emerges from the
black hole's sphere of influence on a timescale $\la 3\times 10^5$~yr.
The shell then stalls unless the black hole mass has reached the value
$M_{\sigma}$ implied by the $M - \sigma$ relation. If the wind shock
did not cool, as suggested here, the resulting (`energy--driven')
outflow would imply a far smaller SMBH mass than actually observed.
In galaxies with large bulges the black hole must grow somewhat beyond
this value, suggesting that the observed $M -\sigma$ relation may
curve upwards at large $M$. Galaxy--wide outflows with velocities
significantly exceeding $\sigma$ probably result from minor accretion
episodes with low gas fractions. These can appear as fossil outflows
in galaxies where there is little current AGN activity. In rare cases
it may be possible to observe the energy--driven outflows which sweep
gas out of the galaxy and establish the black hole--bulge mass
relation. However these require the quasar to be at the Eddington
luminosity.

\section{Acknowledgments}

I thank Ken Pounds and Sergei Nayakshin for illuminating
discussions. Theoretical astrophysics research at Leicester is
supported by an STFC rolling grant.

\end{document}